\documentstyle{aipproc}

\begin{document}
\title{Red Hole Gamma-Ray Bursts: \\
A New Gravitational Collapse Paradigm\\
Explains the Peak Energy Distribution\\ 
And Solves the GRB Energy Crisis
}

\author{James S. Graber$^*$}
\address{$^*$407 Seward Square SE\\
Washington, DC 20003}

%\lefthead{LEFT head}
%\righthead{RIGHT head}
\maketitle

\begin{abstract}
Gamma-ray bursts (GRBs) are still an enigma. In particular the central engine,
 the total energy, and the very narrow distribution of peak energies challenge
 model builders. Motivated by recent theoretical developments (string theory,
 quantum gravity, critical collapse), which suggest that complete gravitational 
collapse can occur without singularities or event horizons, we explore how 
red-hole models (which lack singularities or event horizons) can solve these 
problems better than black-hole models.
\end{abstract}

\section*{KEY GRB MODEL BUILDING CHALLENGES}
Gamma-ray bursts vary rapidly and therefore they must be compact.  Because these 
compact gamma-ray 
bursts release 
enormous energy, they must form an intense fireball that is optically thick, 
pair-producing, 
and thermalized.  But the spectrum is not thermal, and there is no sign of 
pair-production attenuation 
at the high end of the observed spectrum\cite{jsgband}.  This seeming self-contradiction 
(the opacity problem) can be solved by having 
the fireball power a relativistic shell or jet that collides with something 
(perhaps itself) to produce 
the observed gamma rays\cite{jsgrees}.  This fireball/shock model is  
currently the leading candidate 
to explain GRBs\cite{jsgpiran}. It has already overcome several severe model-building 
challenges.  
But like 
almost all other published models, it fails to explain the observed spectroscopy of  
GRBs, particularly 
the narrowness of the observed peak energy distribution\cite{jsgpreece,jsgbrainerd}.  Furthermore, this model does not 
explain the high ratio of the energy of the GRB burst itself (caused by internal shocks)  
to the energy in the afterglow (caused by external 
shocks in the fireball/shock model)\cite{jsgpa}.  
Nevertheless the predictions of this model for the afterglows themselves are consistent with 
current observations\cite{jsgpiran}. 

Finally, there is the problem of the overall energetics of the GRB.  The two leading 
candidates to 
produce the initial fireball or fireballs --the so-called central engine-- are merging 
neutron stars and 
core-collapse supernovae\cite{jsgeichler,jsgwoosley}.  Both these sources have 
over 10$^5$$^4$ ergs of total energy 
available. 
This is more than enough energy for even the most energetic GRB, but it is not at all 
clear how to prevent most of it from 
falling into the 
newly created black hole which forms in the standard general relativity versions 
of these models.

There seems to be an inherent conflict between solving the opacity 
problem and solving the 
peak energy distribution problem.  The only successful technique available to solve 
the transparency problem is to invoke 
highly relativistic bulk motion.  In the relativistic frame, the gamma rays are 
below pair-production threshold and so do not 
suffer pair-production attenuation. This definitively solves the opacity problem.  
But unless the Lorentz gamma factor of 
the bulk motion can be fine-tuned to a very narrow range for all GRBs,  
the resulting blueshift will not only relocate the 
peak of the photon energy distribution; it will also substantially widen it, 
inconsistent with the observed narrow 
E-peak distribution.
Thus one needs to find a way to fine-tune the Lorentz gamma factor or find some 
other way around this conflict.  
In the fireball/shock model the gamma factor depends sensitively on the 
baryon loading, and hence will 
vary widely.  Furthermore, the internal shocks model is dependent on shocks 
with varying Lorentz 
gamma factors colliding with each 
other.  So fine-tuning is not a reasonable option for this model.

 A generic solution to this problem is provided if the relativistic bulk motion results 
not from an initial explosion, 
but rather from the gravitational acceleration of matter falling into a deep potential well.  
An arbitrarily high 
Lorentz gamma factor can be attained, but the accompanying blueshift will be exactly cancelled 
when the matter 
and radiation are redshifted as they emerge from the potential well.  (By that time, the 
matter and radiation 
will have separated, so the opacity problem has already been solved).

A black hole can provide the necessary deep potential well. But once matter or 
radiation is deep in the 
potential well of a 
black hole, it is almost impossible for it to escape.  Therefore, we will consider 
an alternative 
gravitational 
collapse paradigm in which it is possible to escape from deep within the potential well 
of a gravitationally 
collapsed object. 

\section*{WHY CONSIDER ALTERNATE GRAVITY MODELS?}

The problems with constructing a GRB model might be sufficient motivation to consider 
alternate 
theories of gravity.  However, a stronger motivation comes from the theory of gravitation.  
Recent 
theoretical developments in string 
theory, quantum gravity and critical collapse strongly suggest the possibilities of 
both gravitational collapse 
without singularities (and without loss of information) and also gravitational 
collapse without event 
horizons\cite{jsgsv,jsgcm,jsgms,jsgst,jsgchop,jsgchr}.  If these 
possibilities are correct, we are forced to consider the phenomenological consequences 
(such as 
different models for GRBs and core-collapse supernovae) of alternate paradigms for 
gravitational 
collapse in which black holes do not form\cite{jsggrab}.

\section*{RED HOLES-- A NEW PARADIGM}

Many authors have considered the alternative in which a hard core collapsed object similar 
to a smaller 
harder denser neutron star forms in place of a black hole\cite{jsgrob}.   We here consider the 
alternative in 
which no such hard surface forms.  Instead the spacetime stretching that forms a black hole 
in the standard model occurs, but it does not continue to the extent necessary to form an event 
horizon or a singularity.  Instead, spacetime stretches enormously, but not infinitely, and 
forms 
a deep wide potential well with a narrow throat.  We call this a red hole.

This type of spacetime configuration was previously considered by 
Harrison, Thorne, Wakano and Wheeler (HTWW) 
in 1965, but only as a way station in the 
final collapse to a black hole (not yet then called by that name)\cite{jsghtww}.  
In their version, part of 
the 
configuration is inside the event horizon, the collapse continues, and a singularity soon 
forms.  

In the new alternate paradigm we call a red hole, no event horizon forms and no singularity 
forms. The gravitational collapse does not continue forever, but eventually stops.  (Why? 
Perhaps due to quantum effects or string-theory dualities, but we cannot discuss this 
adequately here.)  As the collapse proceeds, the collapsing matter becomes 
denser and denser until it reaches a critical point, after which, the distortion of spacetime 
is so great  that the density decreases.  This happens because the spacetime is stretching 
faster than the collapsing material can fall inward.  (This decreasing density 
effect was already noticed 
by HTWW in their analysis of gravitational collapse 
in the context of standard general relativity\cite{jsghtww}. In general relativity, 
this expansion 
of spacetime is mostly hidden behind the event horizon and does not prevent the formation of 
a singularity in a finite time. This is not the case in several observationally viable alternate 
theories of gravity\cite{jsgrosen,jsgyil,jsgitin}.)  
This is why we are confident that the center of a red hole resembles 
a low-density vacuum more than it resembles a high-density neutron star.  The decrease in 
density due to 
this 
enormous stretching may also be a factor in halting the gravitational collapse of the red hole 
before the stretching becomes infinite.

As a result, even though the stretching of spacetime is enormous, it never becomes fast 
enough to exceed the speed of light and cause an event horizon to form.  It stops 
before it reaches an infinite size or any other form of singularity.  (Infinite density and 
infinite curvature also do not occur.)
Nevertheless, it is very hard to escape from a red hole. First, there are trapped 
orbits inside the red hole for photons as well as massive particles, which allows permanent 
or nearly permanent trapping of mass and energy.  Second, the Shapiro delay in 
crossing a red hole is very substantial, (in some cases, enormous)\cite{jsgshap}. Hence 
particles that are 
only crossing the red hole or passing through are in effect "temporarily" trapped.

In fact most of the matter falling into a red hole will be trapped.  However, radiation, 
and highly 
relativistic matter that falls directly into the center of the red hole and does not rescatter 
while 
inside the red hole, can travel straight through and emerge on the other side.  This 
possibility 
is essential for our proposed new GRB models. 

\section*{RED-HOLE MODELS FOR GRBs}

In order to describe our new red-hole models for GRB's, which are based on modifying the 
existing fireball/shock model, we begin by resummarizing that model.  In the 
fireball/shock model, some form of gravitational collapse deposits a large amount of energy 
in a very small region, (which is called the fireball, and also the central engine). 
The fireball has 
so much energy in such a small space that a 
relativistic expansion must occur.  Part or all of this explosive expansion travels 
through a region 
with a
very small critical number of baryons, which absorb essentially all of the energy and 
form a relativistic blast wave (either spherical or jetted).  Multiple such relativistic shells 
(travelling in the same direction) are created by the central engine, perhaps by repeated 
explosions (possibly due to repeated accretion events).  The faster relativistic shells 
overtake the slower relativistic shells and collide 
with them.  The internal shocks convert the energy of the baryons to gamma rays,(by 
synchroton emission or inverse compton scattering or perhaps by both means). The shells 
eventually 
collide with external matter and generate the main afterglow. (Perhaps an early prompt 
afterglow is 
the result of a reverse shock)\cite{jsgpiran}.  

Basically there are three important sites in this model. First, there is the central engine, or 
fireball site.  (In the standard black-hole interpretation, this is probably near a newly 
forming black hole, perhaps at the pole of a Kerr black hole)\cite{jsgkerr}.  Second, there is the 
location of the internal shocks, where the main gamma-ray burst is generated.  According 
to Piran, this is typically 10$^1$$^2$-10$^1$$^4$ centimeters, or 30-3000 light seconds 
down stream from the central 
engine\cite{jsgpiran}. Third, 
there 
is the location of the external shock, where the relativistic matter collides with material that 
was not part of the original explosion, and the long-lasting (days to months), but 
weak (less total energy than the gamma rays) afterglow is generated.  In the standard model, 
this is far from the central engine.

In  our alternate red hole models we will relocate these three sites in or near a red hole 
instead of near the outside of a black hole. 

In the first and most conservative red-hole model, we merely replace the black hole of 
the standard 
model with a red hole.  The red hole can help the central engine by generating more energy 
than the corresponding black hole 
or by focusing the outgoing jet more narrowly, but the rest of the model is essentially 
the same 
as the standard one and there is no significant impact on the spectral issues.  In other 
words, this first red-hole  
model can help solve the energy crisis, but does not help explain the broad spectrum, with its 
unusual slopes and narrowly distributed peak energy.

In the second -- and  more interesting --  red-hole model, the central 
engine is 
located at the infalling bottleneck of the red hole, and the internal shocks that generate the 
primary gamma-ray burst are located at the outgoing bottleneck of the red hole (which is 
essentially the same place, but at a later time), and the external shocks and the afterglow 
still occur far away at the point where the ejecta encounter the interstellar material  
or some other external 
matter.

In this model, the great internal expansion of the red hole, along with the great acceleration 
of the gravitational infall, help to generate the relativistic jet that will later create 
the GRB.  
Then the focusing effects of the emerging bottleneck of the red hole help to create the 
internal 
shocks necessary for the final transformation of the energy into gamma rays, and to very 
substantially increase the efficiency of this process.  Furthermore, since the blueshift of 
the infall should be exactly cancelled by the redshift of the outclimb, the gamma rays seen 
by the observer will have no net red or blue shift (on average). Therefore the observed 
peak energy will 
be the same as the initial peak energy.  Even if the internal transit involves 
enormous and substantially varying 
Lorentz gamma factors, 
they will not be observed as a net blueshift. So this model helps solve the narrow 
peak energy distribution problem, as well as the energy crisis. It can also help solve the 
spectral wideness and slope problems because 
of the tolerance for differing Lorentz gamma factors during the transit through the red hole.

\end{document}